\documentclass[%
 reprint,
 amsmath,amssymb,
 aps,
prd,
]{revtex4-2}

\usepackage{graphicx}
\usepackage{dcolumn}
\usepackage{bm}
\usepackage{hyperref}
\usepackage[mathlines]{lineno}
\usepackage{xcolor}
\usepackage{soul}

\def\d{\mathrm{d}}

\newcommand{\lra}[1]{{\left\langle #1 \right\rangle}}
\newcommand{\lrp}[1]{{\left( #1 \right)}}
\newcommand{\lrs}[1]{{\left[ #1 \right]}}
\newcommand{\lrc}[1]{{\left\{ #1 \right\}}}

\DeclareMathOperator\erf{erf}

\begin{document}

\preprint{APS/123-QED}

\title{Primordial black holes with mass ratio-modulated initial clustering:\\merger suppression and projected constraints}

\author{Gabriel Luis Dizon}
 \affiliation{National Institute of Physics, University of the Philippines - Diliman}

\date{\today}

\begin{abstract}
We present a modification of the expected local primordial black hole (PBH) count $N(y)$, typically seen in the context of the early PBH binary merger rate as a term in the merger rate suppression. We utilize recent results in small-scale PBH clustering to formulate $N(y)$ in such a way that accounts for variations in the binary mass ratio $q$. We then examine how this change affects the projected constraints on PBH abundance from simulated Einstein Telescope (ET) and LISA mergers. Our results indicate that for broadly extended mass distributions, the merger suppression is greatly reduced for binaries with $q \gg 1$. This leads to an enhanced merger rate for binary distributions favoring a lighter average mass. This change is best reflected by an extension of the low mass end of the constraint windows derived from the resolvable merger channel, although this result is present as well in the stochastic gravitational wave background (SGWB) constraints. Our results imply that the assumption of PBH clustering is testable at abundances and mass ranges much lower than anticipated. Note however that we have only considered the two-body merger channel in this work; a more thorough analysis of our scenario will require a study on the dynamics involved in the three-body merger channel for broad mass distributions, which we leave to future work.
\end{abstract}

\maketitle


\section{\label{sec:intro}Introduction}
Gravitational wave (GW) observations provide one of the few definitive avenues to either confirming or further constraining the primordial black hole (PBH) scenario. We typically study PBHs in this context by analyzing the possibility of PBH binary mergers, and whether these events are detectable by our current and future detectors. Existing studies involving LIGO-Virgo-KAGRA (LVK) observations~\cite{PhysRevD.103.023026, Hutsi_2021, Luca_2020, PhysRevD.110.023040} have a constraint window between $0.1-10^3\,M_\odot$, with the most stringent constraint so far possibly ruling out $O(10\,M_\odot)$ PBHs at abundances higher than $f_\mathrm{PBH} \sim 10^{-4}$~\cite{PhysRevD.110.023040}. Resolvable low-redshift ($z < 30$) detections from future detectors such as Einstein Telescope (ET) and LISA may further constrain the PBH abundance~\cite{PhysRevD.104.083521,De_Luca_2021,Dizon_2024} down to $f_\mathrm{PBH} \sim 10^{-5}$ for PBH masses in the order of $10\,M_\odot$, with the potential of high-redshift ($z>30$) detections allowing us to either test or constrain the abundance up to $f_\mathrm{PBH}\sim 10^{-4}$ at both stellar-mass ($M \sim 10\,M_\odot$) and intermediate-mass ($M \gtrsim 10^{3}\,M_\odot$) scales and greater. A non-detection of the stochastic gravitational wave background (SGWB) in either of these detectors may have the biggest impact on PBH constraints, potentially ruling out PBHs down to the mass range of $10^{-8}\,M_\odot$. We must mention however that these studies are limited by certain assumptions, such as the merger rate formulation used~\cite{PhysRevD.96.123523, Raidal2025} assuming an initial spatial Poisson distribution, or the mass functions assumed are relatively narrow in width. Of course, there have been some attempts to explore departures from these model assumptions, such as incorporating simple non-Gaussian clustering~\cite{De_Luca_2021, Raidal_2017} or broadly extending the mass distribution~\cite{Dizon_2024}. Here we will attempt to explore the possibility of both initial clustering and extended mass distributions affecting GW-derived constraints.

Now, whether PBHs can actually exhibit initial spatial clustering at merger-relevant scales is subject to some discussion within the literature. Here we define "spatial clustering" as the existence of a non-zero two-point PBH correlation function $\xi(r)$, with "initial" meaning that the clustering behavior is already present in the PBH population upon formation. Early studies arguing against the relevance of initial clustering at small scales show that PBHs are not born clustered beyond Poisson at small scales~\cite{PhysRevLett.121.081304}, and that even with broad spectra the relevant correlation distances are longer than the average Poissonian distance~\cite{Dizgah_2019, DELUCA2020135550}, rendering small-scale clustering irrelevant. The latter is particularly interesting, as there are contrasting studies which project that clustering may be relevant for PBH mergers arising from extended mass distributions~\cite{Ballesteros_2018, PhysRevD.98.123533}, particularly with an expectation that there may be an enhancement of the merger rate of light PBH binaries. Quite recently, it has been found through CMB distortion constraint arguments that initial clustering is irrelevant for mergers in the LVK detection window~\cite{Crescimbeni_2025}. However, as the same authors in Ref.~\cite{Crescimbeni_2025} point out, their result may not be extendable to future generation GW detectors, which can probe further into the subsolar and asteroid-mass PBH range. Thus, a good next step forward would be to investigate just that: whether clustering is relevant or even testable for future detectors such as ET and LISA. However, we will utilize a different small-scale clustering model, which not only incorporates exclusion effects into its formulation, but also a way to encode clustering for differing mass ratios~\cite{PhysRevD.109.123538}.

This paper is structured as follows. We sketch the derivation of the expected local PBH count $N(y)$ in Section~\ref{sec:neighbor_count} for both the Poissonian (Section~\ref{subsec:unclustered}) and correlated (Section~\ref{subsec:clustered}) cases. We then present the effects of changing the formulation of $N(y)$ on both the merger rate suppression and the projected $f_\mathrm{PBH}$ constraints from ET and LISA in Section~\ref{sec:results}, borrowing from the methods employed in Ref.~\cite{Dizon_2024} for broadly extended mass distributions. We then state our conclusions and the limitations of our approach at Section~\ref{sec:conclusions}. This work is primarily concerned with the effect \textit{initial} clustering has constraints from PBH mergers. As such, we will be ignoring the effects late-time clustering and DM substructures may have on mergers, although we will briefly discuss at the end of this paper possible implications if these effects are not ignored. We also focus on PBHs and binaries forming in the radiation-dominated era, so initialized quantities such as distance (mass) scales and the merger rate initial conditions will be based off parameters relevant to that era.

\section{\label{sec:neighbor_count}Derivation of local count $N(y)$}
The key quantity encoding first-order clustering behavior is the two-point correlator $\xi(r)$, which is generally defined as~\cite{peebles2020large, Ballesteros_2018, PhysRevLett.121.081304}
\begin{equation}
    1 + \xi(r) = \frac{p(\delta(r) > \delta_c\,|\, \delta(0) > \delta_c)}{p(\delta(0) > \delta_c)}.
    \label{eq:base_correlator}
\end{equation}
$\delta(x)$ here refers to the density contrast at some point $x$, with $\delta_c$ as the critical density threshold for (in our case) PBH formation. The above equation reads as follows: the excess probability over random of two fluctuations separated by comoving distance $r$ to go above threshold $\delta_c$, represented by $1 + \xi(r)$, is simply the conditional probability for some density contrast $\delta(r)$ at a point $r$ from the origin to exceed the threshold, provided that the density contrast at the origin also exceeded the threshold. If $\xi(r) = 0$ (uncorrelated), then this quantity is equal to unity, suggesting that all sites have equal probability of crossing the threshold. Conversely, if Eq.~\ref{eq:base_correlator} deviates from unity, then there is a non-zero correlation between points. This form of correlation is sometimes referred to as the Poisson model~\cite{peebles2020large}.

In the PBH merger context, what primarily matters is not the correlation function itself, but how it affects the expected local count around the PBH binary $N(y)$. In simple terms, the expected number of PBHs enclosed in a spherical volume with comoving radius $R$ is given by
\begin{equation}
    N(R) = \int_0^R\,n\,\d^3r = nV(R),
\end{equation}
where $n$ is the average PBH number density (which we take on average as a constant in space) and $V(R) = 4\pi R^3/3$ is the comoving volume. Given a PBH binary with masses $M_1$ and $M_2$ at some initial distance $x_0$ from each other, the expected number of neighbors at a distance $y > x_0$ from the binary can be estimated by $N(y) = nV(y)$. Since initial clustering is encoded as an excess probability of formation within a defined neighborhood, its effects are naturally expressed in the expected local PBH count $N(y)$. We express this change through the following modification~\cite{peebles2020large, Ballesteros_2018}:
\begin{equation*}
    N(y) = nV(y) \quad\rightarrow\quad N(y) = 4\pi n\int_0^y[1+\xi(r)]r^2\,\d r.
\end{equation*}
Given that the local count $N(y)$ affects the PBH merger rate via the suppression factor $S$, this modification allows us to directly see the effects of initial clustering on the merger rate and resulting constraints, provided that we specify $1 + \xi(r)$. We will do that in Section~\ref{subsec:clustered}. For now, let us first sketch a derivation for $N(y)$ in the unclustered, purely Poissonian case, as this will give us a hint for our clustered derivation later on.

\subsection{\label{subsec:unclustered}Poissonian case: $\xi(r) = 0$}
Ref.~\cite{Raidal2025} provides an expression for the expected local count for initially Poisson-distributed PBHs, written out as
\begin{equation}
    N(y) = \frac{M_\mathrm{tot}}{\langle M \rangle}\frac{f_\mathrm{PBH}}{f_\mathrm{PBH} + \sigma_M},
    \label{eq:raidal_neighbor}
\end{equation}
where $M_\mathrm{tot} = M_1 + M_2$ is the total mass of the PBH binary with component masses $M_1$ and $M_2$, $\lra{M} = \lrp{\int\d\ln{M}\,\psi(M)}^{-1}$ is the average PBH mass for some underlying mass distribution $\psi(M)$, $f_\mathrm{PBH}$ is the PBH abundance, and $\sigma_M \simeq 3.6\cdot10^{-5}$ is the rescaled variance of early matter perturbations. We provide a derivation here for completeness, following a similar derivation of the merger rate by Ref.~\cite{Raidal_2017}, although keep in mind that our relations here rough estimates of scale.

We can straightforwardly evaluate $N(y) = nV(y)$ for a comoving spherical volume, yielding
\begin{equation}
    N(y) = 4\pi n\int_0^y r^2\,\d r,
    \label{eq:initial_neighbor_integral}
\end{equation}
where we set $\xi(r) = 0$, as we are dealing with purely Poissonian distributions. The characteristic distance scale of a PBH binary with total mass $M_\mathrm{tot}$ can be estimated as~\cite{Raidal_2017, PhysRevD.96.123523, Nakamura_1997, PhysRevD.58.063003}
\begin{equation}
    \tilde{x} = \lrp{\frac{3}{4\pi}\frac{M_\mathrm{tot}}{a_\mathrm{eq}^3\rho_\mathrm{eq}}}^{1/3},
    \label{eq:tilde_x}
\end{equation}
where $a_\mathrm{eq}$ is the scale factor and $\rho_\mathrm{eq}$ is the energy density at matter-radiation equality. Similarly, the average characteristic scale for the PBH population in general is estimated to be
\begin{equation}
    x_\mathrm{ave} = \lrp{\frac{3}{4\pi}\frac{\lra{M}}{a_\mathrm{eq}^3\rho_\mathrm{eq}}}^{1/3}.
    \label{eq:x_ave}
\end{equation}
Given the above, we can write out the average number density $n$ as just the PBH abundance $f_\mathrm{PBH}$ over the comoving volume bounded by the average characteristic scale~\footnote{Ref.~\cite{Ballesteros_2018} defines $n$ through the maximum allowed binary distance at which PBH binaries can still be bound, i.e., $x_\mathrm{max} \equiv (3f_\mathrm{PBH}/4\pi n)^{1/3}$. This treatment seems to be only valid for strictly monochromatic mass distributions.} 
\begin{equation}
    n = \frac{3f_\mathrm{PBH}}{4\pi x_\mathrm{ave}^3}.
    \label{eq:num_density}
\end{equation}
Finally, given how $y$ is set to be the minimum distance scale at which perturbing neighbors can approach the binary without disrupting it, it is reasonable to estimate $y$ to be, at the minimum, of the order of the characteristic scale of the binary
\begin{equation}
    y \sim \tilde{x} = \lrp{\frac{3}{4\pi}\frac{M_\mathrm{tot}}{a_\mathrm{eq}^3\rho_\mathrm{eq}}}^{1/3}.
    \label{eq:y_approx}
\end{equation}
Plugging in Eqs.~\eqref{eq:num_density} and~\eqref{eq:y_approx} into~\eqref{eq:initial_neighbor_integral}, we get
\begin{align}
    N(y) &= 4\pi n\int_0^y r^2\,\d r, \nonumber\\
    &= \frac{f_\mathrm{PBH}}{x_\mathrm{ave}^3}y^3.
\end{align}
We then plug in Eq.~\eqref{eq:x_ave} to obtain
\begin{equation}
    N(y) = \frac{M_\mathrm{tot}}{\lra{M}}f_\mathrm{PBH}.
\end{equation}
This expression is quite similar to Eq.~\eqref{eq:raidal_neighbor}, with the sole exception of the $f_\mathrm{PBH}$ factor. The latter factor in Eq.~\eqref{eq:raidal_neighbor} is expressed as $f_\mathrm{PBH}/(f_\mathrm{PBH} + \sigma_M)$, where $\sigma_M$ is the rescaled variance of matter density perturbations upon binary formation. This formulation arose by considering only initial conditions that produce binaries expected to survive until the collapse of the first DM structures~\cite{Raidal2025}. More intuitively, we say that the $f_\mathrm{PBH}$ factor in Eq.~\ref{eq:raidal_neighbor} accounts for the local abundance of PBHs (through $f_\mathrm{PBH}$), weighted against the total distribution of perturbing bodies (both PBH and other matter perturbations) around the binary (through $f_\mathrm{PBH} + \sigma_M$). We stick with this factor for our modification of $N(y)$ in the next subsection, as we see it as a reasonable way to incorporate the effects of non-PBH neighbors for binaries that form in the radiation-dominated era, even if its impact is minimal.

\subsection{\label{subsec:clustered}Correlated case: $\xi(r) \neq 0$}
Having developed the Poisson case as a baseline, we can now move on to the clustered case. We do have several options to choose from for our desired formulation of $\xi(r)$, be it based off threshold statistics~\cite{Ballesteros_2018, PhysRevLett.121.081304}, excursion set methods~\cite{Dizgah_2019, PhysRevD.98.123533}, or even through the introduction of non-Gaussianities~\cite{Young_2020, Raidal_2017, De_Luca_2021}. We opt for a relatively recent formulation developed through excursion set methods~\cite{PhysRevD.109.123538}, which we utilize due to its capacity to encode the correlation behavior between PBHs of differing masses. The correlation function is defined as
\begin{widetext}
\begin{equation}
    1 + \xi_{\sigma_1, \sigma_2}(r) = \frac{e^{\lambda^2 (w_1 + w_2 - 1)}}{\sqrt{\pi}\lambda}\sqrt{\frac{w_{12}}{(1-w_1w_2)^3}}\left[1 + \lambda\sqrt{\pi w_{12}}\left(\frac{1}{2\lambda^2w_{12}} + 1\right)e^{\lambda^2w_{12}}\erf(\lambda\sqrt{w_{12}})\right],
    \label{eq:correlation_function}
\end{equation}
\end{widetext}
where $\sigma_1$ and $\sigma_2$ define the sizes (or masses) of the PBHs. $\lambda = \delta_c / \sqrt{2\sigma_r}$ defines the absorbing threshold for PBH formation, where $\delta_c$ is the critical density threshold. $\sigma_r$ defines a variance scale between the PBHs with distance $r$ from each other, and $w_i = \sigma_r/\sigma_i$ where $i = \{1,2\}$. We also introduce a shorthand $w_{12} = (1-w_1)(1-w_2)/(1-w_1w_2)$. Apart from allowing for non-symmetric binary masses, this formulation also accounts for exclusion effects due to the assumption that the PBHs are \textit{not} pointlike, in contrast to previous Poissonian formulations. This brings about interesting effects, particularly anti-correlation at short distances.

Now in order for Eq.~\eqref{eq:correlation_function} to be actually useful, we must specify a way to relate the variance $\sigma_r$ to the comoving distance scale $r$. Because $\sigma_r$ also directly relates to the sizes and masses of the PBHs involved, it is typically defined as
\begin{equation}
    \sigma_r = \int_0^\infty \mathcal{P}_\delta(k) W^2(k, r)\,\mathrm{d}\ln k,
    \label{eq:variance_r}
\end{equation}
where $\mathcal{P}_\delta(k)$ is the power spectrum of PBH-forming fluctuations $\delta$, and $W(k,r)$ is the window function relating the fluctuation wavenumbers $k$ with the comoving distance scale $r$. We choose a power spectrum consistent with our choice of mass functions, i.e., power law. Thus a natural choice would be
\begin{equation}
    P_\delta(k; n_p) \sim \lrp{\frac{k}{k_*}}^{n_p},
\end{equation}
where $n_p$ is the spectral tilt, and $k_*$ defines a cutoff scale for the spectrum. Normally this spectrum is then passed through $\sigma_r$ and into the PBH mass fraction $\beta(M)$ in order to generate the corresponding mass distribution. We opt for a less rigorous approach, inspired by the heuristic derivation done by Ref.~\cite{DELUCA2020135550}. Their approach showed that a broad PBH fluctuation power spectrum with flat tilt ($n_p = 0$) may generate unimodal mass distributions with high mass tails scaling as $M^{-3/2}$. If we define our power law mass distributions as $\psi(M) \sim M^{\gamma-1}$, their result would correspond to a power law exponent of $\gamma = -1/2$.  This sets a correspondence between the tilt of the fluctuation power spectrum $n_p$ and the tilt of the power law PBH mass distribution $\gamma$. We may extrapolate this to be a linear relation about the neighborhood of $n_p \simeq 0$, i.e., $n_p = \gamma + 1/2$. We stress however that this is a \textit{heuristic extrapolation} for values of $\gamma \simeq -1/2$; the actual relation between the tilt exponents may not be linear for values of $\gamma$ much smaller or larger than -1/2.

For now, we have a simple and consistent relation between the power spectrum and the mass distribution via their exponents. In order to calculate for $\sigma_r$, we choose a top-hat in $k$-space window function selecting for the modes between $k \in [k_*, 1/r]$. Because we are only interested in the \textit{small-scale} clustering behavior between PBHs, we set the high-mass cutoff scale at $k_* \sim 1/\tilde{x}$, i.e., at the characteristic scale of the PBH binary being examined. Evaluating Eq.~\eqref{eq:variance_r} with these considerations gives us two cases
\begin{equation}
    \sigma_r =
    \begin{cases}
        \frac{1}{n_p}\lrs{\lrp{\frac{\tilde{x}}{r}}^{n_p} - 1}, \quad &n_p \neq 0,\\
        \ln(\tilde{x}/r),\quad &n_p = 0.
    \end{cases}
    \label{eq:variances_real}
\end{equation}
We may then invert the equations above to obtain the corresponding real-space comoving distances
\begin{equation}
    r =
    \begin{cases}
        \tilde{x}\,\lrc{n_p\sigma_r +1}^{-1/n_p}, \quad &n_p\neq 0,\\
        \tilde{x}\,\exp(-\sigma_r),\quad &n_p = 0.
    \end{cases}
\end{equation}

Now, recall that expected local count with non-zero correlations is given by
\begin{equation}
    N(y) = 4\pi n\int_0^y[1+\xi(r)]r^2\,\d r.
    \label{eq:neighbor_mod}
\end{equation}
Having already set relations for $\sigma_r$ and $r$, we calculate for the integral bounds in $\sigma_r$-space as follows. The real-space upper bound $y$ is trivial to express in $\sigma_r$, as $y \sim \tilde{x}$ and we have set our cutoff scale at $\tilde{x}$. Thus $\sigma_\mathrm{max} = 0$ regardless of the value of $\gamma$. A bit more nuance is required, however, for the lower bound. Because our correlation function of choice assumes PBHs have finite non-zero size, we cannot arbitrarily set $r\rightarrow0^+$ (or $\sigma_r\rightarrow+\infty$). Thus we require a finite lower bound, which we set to be at the scale of the larger PBH in the binary. If we assume that $M_2 \geq M_1$, then its corresponding comoving distance scale is given by
\begin{equation}
    r(M_2) = \lrp{\frac{3}{4\pi}\frac{M_2}{a_\mathrm{eq}^3\rho_\mathrm{eq}}}^{1/3}.
\end{equation}
Define the binary mass ratio $q = M_2/M_1 \geq 1$. We can then express the total binary mass as $M_\mathrm{tot} = M_1 + M_2 = (q+1)M_1$. Plugging these values in to Eqs.~\eqref{eq:tilde_x} and~\eqref{eq:variances_real}, we get the lower bound
\begin{equation}
    \sigma_\mathrm{min} = \frac{1}{n_p}\lrs{\lrp{\frac{q + 1}{q}}^{n_p/3} - 1}, 
\end{equation}
for spectral tilts $n_p \neq 0$, and
\begin{equation}
    \sigma_\mathrm{min} = \frac{1}{3}\ln\lrp{\frac{q+1}{q}}
\end{equation}
for $n_p = 0$.

Finally, we note that the differential radial element $r^2\d r$ transforms as
\begin{equation}
    r^2\d r = -\tilde{x}^3\,\d\sigma_r\times
    \begin{cases}
        \lrs{n_p\sigma_r + 1}^{-\frac{n_p + 3}{n_p}}, \quad &n_p\neq 0,\\
        \exp\lrp{-3\sigma_r}, \quad &n_p = 0.
    \end{cases}
\end{equation}
Combining this, our new integration bounds, and Eq.~\eqref{eq:num_density} into Eq.~\eqref{eq:neighbor_mod}, we get our expression for the expected local count with mass-ratio modulated initial clustering
\begin{equation}
N(y, q) = \frac{M_\mathrm{tot}}{\lra{M}}\frac{f_\mathrm{PBH}}{f_\mathrm{PBH} + \sigma_M} \mathcal{G}(q; \delta_c, n_p),
\label{eq:general_neighbor_count}
\end{equation}
where the mass ratio dependence is encoded in
\begin{widetext}
\begin{equation}
    \mathcal{G}(q; \delta_c, n_p) = 3 \int_0^{\sigma_\mathrm{min}(q)}\mathrm{d}\sigma_r\times 
\begin{cases}
    \lrs{n_p\sigma_r + 1}^{-\frac{n_p + 3}{n_p}}\,[1 + \xi_{\sigma_1, \sigma_2}(\sigma_r; \delta_c)], \quad &n_p\neq-0,\\
    \exp\lrp{-3\sigma_r}\,[1 + \xi_{\sigma_1, \sigma_2}(\sigma_r; \delta_c)], \quad &n_p = 0.
\end{cases}
\label{eq:mass_ratio_factor}
\end{equation}
\end{widetext}
Here, we explicitly show the density collapse threshold $\delta_c$ as a parameter of $\xi(r)$. We define $N$ as explicitly a function of both $y$ and $q$ to emphasize that the binary mass ratio enters into the equation through $\sigma_\mathrm{min}$. Note that at the limit of $\xi(r) = 0$ and $\sigma_\mathrm{min} \rightarrow +\infty$, the integral in Eq.~\eqref{eq:general_neighbor_count} reduces to $1/3$ regardless of spectral tilt $n_p$, recovering the pure Poissonian expected local count of Eq.~\eqref{eq:raidal_neighbor}. As we can see then, the effect of mass ratio modulation is simply a mass ratio-dependent rescaling on the Poissonian expected local count: i.e., $N(y, q) = \bar{N}(y)\mathcal{G}(q;\delta_c,n_p)$. Interestingly, if we turn off just the correlation function $\xi(r)$ but retain the exclusion induced by a finite $\sigma_\mathrm{min}$, we find that at the high mass ratio limit, i.e., $q \rightarrow + \infty$, $N(y)$ vanishes. This is expected, as increasing $q$ lets the larger mass $M_2$ approach the total mass of the binary, effectively reducing the possibility of other PBHs from occupying the same volume.

\begin{figure}
    \centering
    \includegraphics[width=\linewidth]{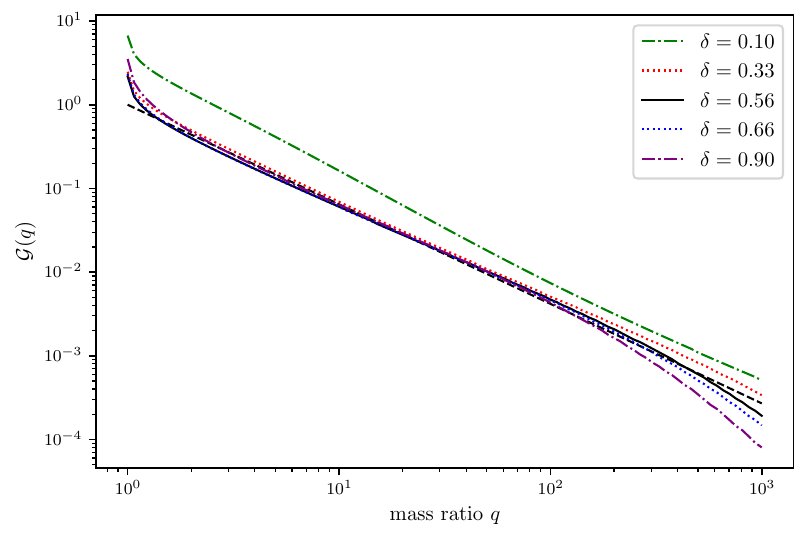}\\

    \includegraphics[width=\linewidth]{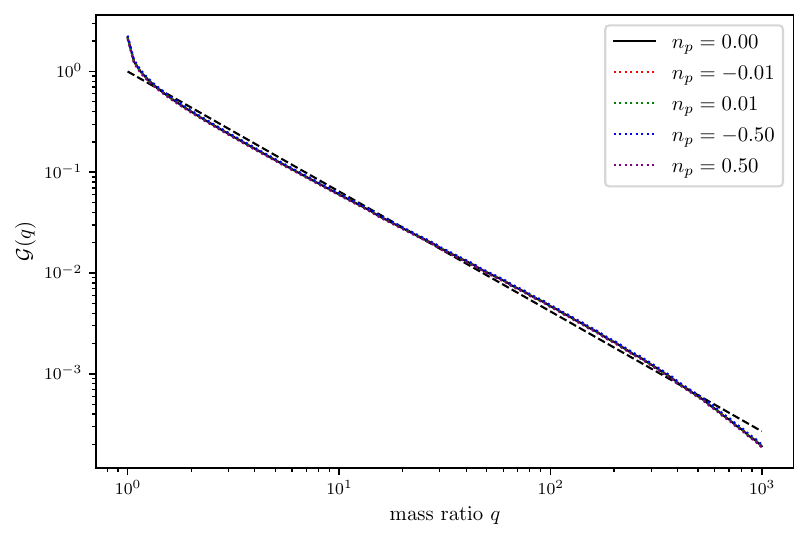}
    \caption{Local count mass ratio dependence $\mathcal{G}(q)$ with varying values of $\delta_c$ (top panel, fixed $n_p = 0$) and varying values of spectral tilt $n_p$ (bottom panel, fixed $\delta_c = 0.56$). In both panels, the black solid line corresponds to the case where $\delta_c = 0.56$ and $\gamma = -1/2$, while the black dashed line is approximately $\mathcal{G}(q; \delta_c=0.56, n_p = 0) \approx q^{-6/5}$. Note that for the top panel, $\delta_c = 0.10, 0.90$ are both represented by dash-dotted lines, while $\delta_c = 0.33, 0.66$ are represented by dotted lines. For the bottom panel, all dotted curves ($n_p = -0.5, \,-0.01,\,0.01,\,0.5$) almost completely overlap with the $n_p = 0$ case.}
    \label{fig:g-q}
\end{figure}

Figure~\ref{fig:g-q} shows how $\mathcal{G}(q)$ changes when we vary threshold $\delta_c$ (top panel) and spectral tilt $n_p$ (bottom panel). In both panels, the black solid line corresponds to the case $\delta_c = 0.56$ and $n_p = 0$, while the black dashed line is a curve-fit approximation of the same function case, where $\mathcal{G}(q; \delta_c=0.56, n_p = 0)$ behaves approximately as $q^{-6/5}$. On the top panel, we see how little variation there is in the behaviour of $\mathcal{G}(q)$ from $\delta_c = 0.33$ (dotted red line) to $\delta_c = 0.66$ (dotted blue line) for fixed $\gamma = -1/2$. This suggests that our form of $\mathcal{G}(q)$ is insensitive to variations in $\delta_c$ within the interval considered in literature~\cite{PhysRevD.100.123524, PhysRevD.101.044022}, at least within the mass ratio interval $2 \lesssim q \lesssim 100$. For values beyond this interval, the behaviour of $\mathcal{G}(q)$ for $\delta = 0.90$ (dash-dotted purple line) deviates from the $\delta_c = 0.56$ case up to at most a fifth of an order of magnitude in the mass ratio interval $100 \lesssim q \leq 1000$. The case of $\delta_c = 0.10$ (dash-dotted green line) exhibits the greatest departure from all the other cases, with values of $\mathcal{G}(q)$ being at most three times greater than the other curves across all mass ratios. A lower threshold may imply a higher $N(y)$ multiplier due to the lower barrier for PBH formation, although it may also indicate a breakdown of the model for low values of $\delta_c$. For our purposes, we maintain a value of $\delta_c = 0.56$, following the treatment by Ref.~\cite{PhysRevD.103.063538} for broad mass functions.

We also plot in the bottom panel how $\mathcal{G}(q)$ changes with the spectral tilt $n_p$ for fixed threshold $\delta_c = 0.56$. We see that the behaviour of $\mathcal{G}(q)$ is unchanging across the range of values $n_p \in [-0.5, 0.5]$. Under our extrapolation ($n_p = \gamma + 1/2$), this range of spectral tilts would correspond to power law tilts $\gamma \in [-1, 0]$. This result suggests that our model of mass ratio modulated clustering is insensitive to the spectral tilt $n_p$ of power law-distributed fluctuation spectra. That said, we must emphasize that this trend may not extend generally to the corresponding mass distribution as suggested by our heuristic ansatz. At best, we can say that $\mathcal{G}(q)$ is insensitive to changes in the neighhborhood of $n_p \simeq 0$, which corresponds to roughly the power law mass distributions with exponent $\gamma \simeq -1/2$. Moving forward, we only consider density threshold $\delta_c = 0.56$ and spectral tilt $n_p = 0$, corresponding to a power law distribution with $\gamma = -1/2$.

\begin{figure}
    \centering
    \includegraphics[width=\linewidth]{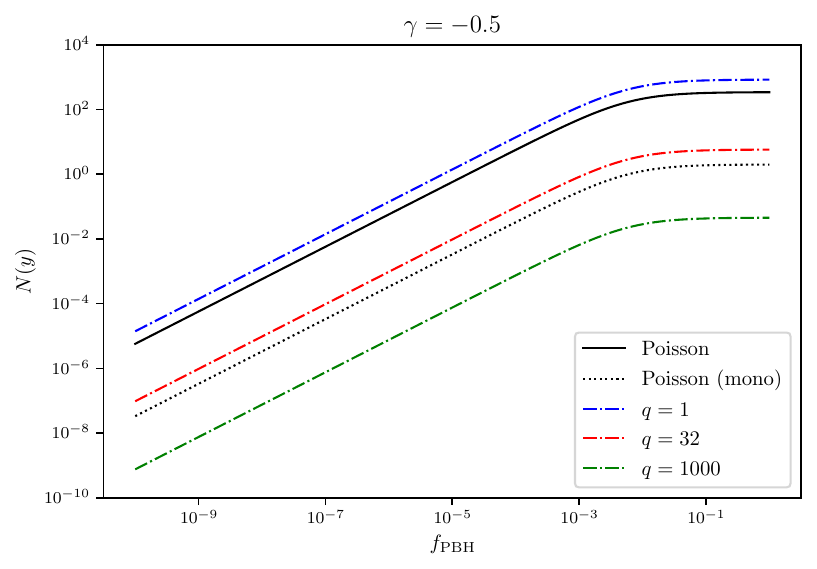}
    \caption{Expected local count $N(y)$ for varying clustering models, plotted against PBH abundance $f_\mathrm{PBH}$ for power law exponent $\gamma = -1/2$. Black lines represent Poisson-distributed PBHs modeled by Eq.~\eqref{eq:raidal_neighbor}, with the black dotted lines showing the monochromatic case and the black solid lines showing the power law case. Colored dash-dotted lines show the clustered case with Eq.~\eqref{eq:general_neighbor_count} as the model, with blue, red, and green representing mass ratios $q = 1, 32, 1000$ respectively. All power law plots assume $M_\mathrm{max} = M_\mathrm{tot} = 30\,M_\odot$ and maximum mass ratio $q_\mathrm{max} = 1000$.}
    \label{fig:neighbor_count}
\end{figure}

We plot in Figure~\ref{fig:neighbor_count} the expected local count $N(y)$ as a function of $f_\mathrm{PBH}$ for power law exponent $\gamma = -1/2$. The black lines correspond to the purely Poisson case, with the dotted black line representing the case for a monochromatic mass function, which approaches $N \approx 2$ for $f_\mathrm{PBH} \geq 10^{-2}$. The colored dash-dotted lines show the expected local count for the correlated case for different values of the binary mass ratio $q$. All the non-monochromatic cases assume a power law distribution where $M_\mathrm{max} = M_\mathrm{tot} = 30\,M_\odot$ and the maximum mass ratio is $q_\mathrm{max} = 1000$. As stated above, we fix $\delta_c = 0.56$, following Ref.~\cite{PhysRevD.103.063538}.

There are two key points to note in the Figure. The first is that across the board, $N(y)$ decreases by orders of magnitude with increasing $q$, although the rate of decrease is not uniform in $q$. This is expected, as $\mathcal{G}(q) \sim q^{-6/5}$ under our parameters, at least within the interval $2 \lesssim q \lesssim 1000$. In particular, we can see around two decades of reduction going from $q = 1$ to $q = 32$, and another two decades going from $q = 32$ to $q = 1000$. We may infer that $\mathcal{G}(q)$ approaches zero for very large mass ratios, i.e., $q \rightarrow +\infty$, although the falloff speed beyond $q \simeq 1000$ may not necessarily go as $q^{-6/5}$. 

The second key point is that mass ratios in the range of $q \sim O(1)$ actually yield higher values of $N(y)$ in the correlated case compared to their purely Poissonian counterparts. This is particularly important, as slight changes in the expected local count can significantly affect the behavior of the initial suppression with respect to the PBH abundance. As we will show in the next section, this split modulation of $N(y)$ through the binary mass ratio $q$ has consequences not only in the differential PBH binary merger rate, but also in the detection likelihood of both resolvable and SGWB mergers, as reflected in the PBH abundance constraints later. 

\section{\label{sec:results}Application to PBH binary mergers}
With a set of expressions for the expected local count $N(y)$ for both Poissonian and correlated cases, we are ready to explore the effects non-zero spatial correlation have on PBH binary mergers and the resulting constraints on the PBH abundance. We adopt the merger rate in Ref.~\cite{Raidal2025}, written out as
\begin{align}
&\frac{\mathrm{d}^2R_\mathrm{PBH}}{\mathrm{d}M_1\mathrm{d}M_2} = \frac{1.6\times 10^6}{\mathrm{Gpc}^3\,\mathrm{yr}}f^{53/37}_\mathrm{PBH}\eta^{-34/37}\left(\frac{t}{t_0}\right)^{-34/37} \times \nonumber\\
&\times\left(\frac{M_\mathrm{tot}}{M_\odot}\right)^{-32/37} S_1(M_1, M_2, f_\mathrm{PBH})\psi(M_1)\psi(M_2),
\label{eq:merger_rate}
\end{align}
where $\eta = q/(q+1)^2$ is the symmetric mass ratio, $t$ is the merger time, $t_0$ is the current age of the Universe, $S_1$ is the initial merger rate suppression factor, and $\psi(M)$ is the PBH mass distribution. Following Ref.~\cite{Dizon_2024}, we assume $\psi(M)$ to have the form of a double-bounded power law written out as
\begin{equation}
    \label{eq:pl}
    \psi_{\mathrm{PL}}(M; \gamma) =
    \begin{cases}
        \mathcal{N}_\mathrm{PL}M^{\gamma - 1}, &\, M \in [M_\mathrm{min}, M_\mathrm{max}], \\
        0, &\, \mathrm{otherwise,}
    \end{cases}
\end{equation}
where $\mathcal{N}_\mathrm{PL}$ is the normalization factor given by
\begin{equation}
    \mathcal{N}_\mathrm{PL} = 
    \begin{cases}
        \frac{\gamma}{M_\mathrm{max}^\gamma - M_\mathrm{min}^\gamma}, &\quad \gamma \neq 0 \\
        \frac{1}{\log(M_\mathrm{max}/M_\mathrm{min})}, &\quad \gamma = 0.
    \end{cases}
\end{equation}
We set a maximum allowed mass ratio $q_\mathrm{max} = M_\mathrm{max}/M_\mathrm{min} = 1000$. We also calculate for the $i$-th moments of the mass distribution through
\begin{equation}
\label{eq:expectation_value}
    \langle M^i \rangle = \frac{\int\,\d\ln{M}\,\psi(M)M^i}{\int\,\d\ln{M}\,\psi(M)}.
\end{equation}

For the initial merger rate suppression, we utilize an approximate form given by~\cite{Hutsi_2021} 
\begin{equation}
\label{eq:suppression_1}
    S_1 \approx 1.42\left[\frac{\lra{M^2}/\lra{M}^2}{N(y) + C} + \frac{\sigma^2_M}{f^2_\mathrm{PBH}}\right]^{-21/74}e^{-N(y)}, 
\end{equation}
where $C$ is a function in $f_\mathrm{PBH}$ with the form
\begin{align}
    &C(f_\mathrm{PBH}) = f_\mathrm{PBH}^2\frac{\lra{M^2}/\lra{M}^2}{\sigma_M^2}\times\nonumber\\
    &\times\left\{\left[\frac{\Gamma(29/37)}{\sqrt{\pi}}U\left(\frac{21}{74},\frac{1}{2},\frac{5f_\mathrm{PBH}^2}{6\sigma^2_M}\right)\right]^{-74/21} - 1\right\}^{-1}.
    \label{eq:cfit}
\end{align}
$U(a, b; z)$ is the confluent hypergeometric function. We must note that this approximation for the suppression has varying levels of deviation from its exact formulation in Ref.~\cite{Raidal2025} for the double-bounded mass distributions we consider above. In particular, Ref.~\cite{Dizon_2024} reports deviations up to $50\%$ values of $\gamma \gg 1$. Given that we are interested in flat mass distributions, i.e., $\gamma \simeq 0$, we can safely sidestep this issue. Also, as stated in the Section~\ref{sec:intro}. we are only interested in the effects of \textit{initial} clustering, and hence, initial suppression. Thus we ignore the time-dependent $S_2$ suppression term, which accounts for disruption from later-forming DM haloes and substructures~\cite{PhysRevD.101.043015}. We will briefly discuss in Section~\ref{sec:conclusions} the implications of our results on late-time suppression, as well as possible ways to model such behaviour.

\subsection{\label{subsec:suppression}Effect on initial suppression}

\begin{figure}
    \centering
    \includegraphics[width=\linewidth]{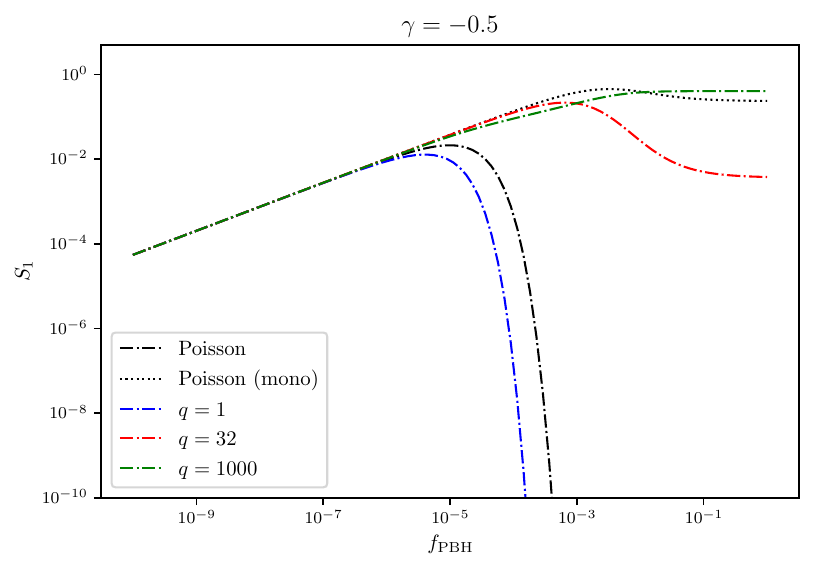}
    \caption{Initial merger suppression $S_1$ for varying clustering models, plotted against PBH abundance $f_\mathrm{PBH}$ for power law exponent $\gamma = -1/2$. Black lines represent Poisson-distributed PBHs modeled by Eq.~\eqref{eq:raidal_neighbor}, with the dotted and dash-dotted lines showing the monochromatic and power law cases respectively. Colored dash-dotted lines show the clustered case with Eq.~\eqref{eq:general_neighbor_count} as the model, with blue, red, and green representing mass ratios $q = 1, 32, 1000$ respectively. All non-monochromatic plots assume power law distributions where $M_\mathrm{max} = M_\mathrm{tot} = 30\,M_\odot$ and maximum mass ratio $q_\mathrm{max} = 1000$.}
    \label{fig:suppression}
\end{figure}

Figure~\ref{fig:suppression} shows the value of the initial merger suppression $S_1$ as a function of $f_\mathrm{PBH}$ for $\gamma = -1/2$. The black lines correspond to the purely Poisson case, with the dotted and dash-dotted black lines representing the case for a monochromatic and a power law mass distribution respectively. The colored dash-dotted lines show the suppression trend for the correlated case for different values of the binary mass ratio $q$. All the non-monochromatic cases assume power law distributions where $M_\mathrm{max} = M_\mathrm{tot} = 30\,M_\odot$ and the maximum mass ratio is $q_\mathrm{max} = 1000$. Just as for Figure~\ref{fig:neighbor_count}, we fix the collapse threshold as $\delta_c = 0.56$. We utilize this value of the collapse threshold for all subsequent subsections and figures.

Here we can clearly see how the suppression term $S_1$ behaves as change the value of the PBH abundance. At $f_\mathrm{PBH} < 10^{-5}$, the suppression values are the same across all models and mass ratios, increasing log-linearly with the abundance. Above this value, we start to see changes in the suppression behavior for each model. In particular, the $q = 1$ correlated binary merger channel is quickly suppressed at $10^{-4} \lesssim f_\mathrm{PBH} \lesssim 10^{-3}$, followed shortly by the Poisson power law case. Within this same region, we see the the $q=32,\,1000$, and monochromatic cases still increase with $f_\mathrm{PBH}$. The $q = 32$ channel reaches its maximum at $f_\mathrm{PBH} \sim 10^{-3}$, experiencing slight suppression at higher values of abundance before settling at a value. We thus see a consistent trend: the monochromatic and $q = 1000$ binary channels are the least suppressed, and the $q = 1$ and Poisson power law binary channels are the most suppressed.

Before we move on to the constraints, it is important that we properly contextualize these plots in Figure~\ref{fig:suppression}. These suppression plots assume a power law distribution where its maximum allowed mass $M_\mathrm{max}$ is also the total binary mass $M_\mathrm{tot}$ considered. Now recall that all formulations of $N(y)$ we consider have a factor of $M_\mathrm{tot}/\lra{M}$, where $\lra{M}$ is the average mass of the distribution. By setting $M_\mathrm{tot} = M_\mathrm{max}$, we have maximized this factor for each distribution we show in Figure~\ref{fig:suppression}, effectively giving us the maximum possible suppression for some combination of $f_\mathrm{PBH}$ and $q$. We can reduce the suppression for the worst case scenarios (i.e., $q \gtrsim 1$) by shifting down the total binary mass $M_\mathrm{tot}$ to be closer to the average mass $\lra{M}$. However, because the mass factor is shared across all formulations of $N(y)$, this is rescaling applies across all $N(y)$ models, which means mergers from symmetric and nearly symmetric binaries in the correlated case are \textit{always} going to be suppressed sooner and harder at higher abundances than the case where binaries have neighbors that are purely Poisson-distributed. Shifting down $M_\mathrm{tot}$ while freezing the value of $\lra{M}$ also presents another trade-off, as keeping the average mass constant also effectively freezes the mass distribution $\psi(M)$. Smaller values of $M_\mathrm{tot}$ thus have a more limited range of possible mass ratios accessible to them, due to being closer to the lower mass cutoff $M_\mathrm{min}$. 

While these plots show correlated $q \sim O(1)$ binaries are over-suppressed relative to the Poisson power law case, we must remind ourselves that for the latter case, the suppression shown here applies to \textit{all} binaries in the distribution with the same total mass $M_\mathrm{tot}$. This is regardless of mass ratio $q$. Thus if we compare the two Poisson cases, monochromatic and power law, we see that binaries heavier than average sampled from power law distributions are always more suppressed than their monochromatic counterparts no matter the ratio of the binary component masses. Our formulation of $N(y)$ in Eq.~\eqref{eq:general_neighbor_count} changes this, taking advantage of combining finite-size exclusion effects with short range \textit{anti-correlation} between highly asymmetric binaries to reduce the likelihood of merger suppression between the two. At the PBH abundance range of $f_\mathrm{PBH} \geq 10^{-4}$, this may lead to an enhancement of the merger rate despite the over-suppression of the $q \gtrsim 1$ binary channels. Despite this, we will see in the next subsection that enhancement in the merger rate may not always translate to a wider testable constraint window. 

\subsection{\label{subsec:constraints}Effect on projected GW constraints}
Knowing now how our formulation of $N(y)$ affects the initial merger rate suppression, we may now proceed to compute for the differential merger rate and projected constraints that may be tested by future detectors. Before we continue, let us first list down the constraints from other observations that we will be comparing our results to. 

Figures~\ref{fig:resolvable} and~\ref{fig:SGWB} show the projected constraints from resolvable PBH mergers and SGWB-generating PBH mergers, respectively. The gray constraint curves each correspond to an observation \textit{ruling out} that combination of $f_\mathrm{PBH}$ and $\lra{M}$ parameters from consideration. We include here microlensing observations from Subaru HSC~\cite{Niikura2019,PhysRevD.101.063005}, MACHO/EROS (E)~\cite{Alcock_2001_1,Alcock_2001}, OGLE (O)~\cite{PhysRevD.99.083503}, and Icarus (I)~\cite{PhysRevD.97.023518}. We also consider constraints from Lyman~$\alpha$ (L-$\alpha$) observations~\cite{PhysRevLett.123.071102}, x-ray binary (XRB) observations~\cite{Inoue_2017}, as well as dynamical constraints from dwarf galaxy dynamics (Seg I)~\cite{PhysRevLett.119.221104}, dwarf galaxy heating (DGH)~\cite{Lu_2021, Takhistov_2022}, and dynamical friction (DF)~\cite{Carr_1999}. On the higher mass end, we include PBH accretion constraints from Planck observations~\cite{PhysRevResearch.2.023204}, the CMB $\mu$-distortion ($\mu$-dist.)~\cite{PhysRevD.97.043525}, and from estimates of the neutron-to-proton (n/p) ratio~\cite{PhysRevD.94.043527}. Finally, we also include the resolvable merger constraint from the O3 run of the LVK detector (GWO3)~\cite{PhysRevD.110.023040}. Note that for the high-redshift case in Figure~\ref{fig:resolvable}, we chose to omit the microlensing, dynamical, XRB, and GWO3 constraints. We do not expect these observations to have an impact at these higher redshift ranges.

\subsubsection{Resolvable mergers}
The main observable considered when setting constraints from resolvable mergers is the expected number of merger events per year $N_\mathrm{det}$, given by~\cite{Dominik_2015, Raidal2025}
\begin{align}
N_\mathrm{det} = \int\mathrm{d}z\,\mathrm{d}M_1\mathrm{d}M_2\,\frac{1}{1+z}\frac{\mathrm{d}V_c(z)}{\mathrm{d}z}\times\nonumber\\
\times\frac{\mathrm{d}^2R_\mathrm{PBH}}{\mathrm{d}M_1\mathrm{d}M_2}p_\mathrm{det}(M_1, M_2, z).
\label{eq:ndet}
\end{align}
$V_c(z)$ is the comoving volume per unit redshift, and $p_\mathrm{det}$ is the binary detectability of any mass pair $M_1$ and $M_2$ at a redshift $z$ for some particular GW detector, and $R_\mathrm{PBH}$ is the differential PBH merger rate given by Eq.~\ref{eq:merger_rate}. $p_\mathrm{det}$ is determined by the ratio between the optimal detector signal-to-noise ratio (SNR) $\rho_\mathrm{det}$ and a preset detection threshold $\rho_\mathrm{thr} = 8$. We simulate the detector SNR data $\rho_\mathrm{det}$ using the Python package \texttt{gwent}~\cite{Kaiser:2021}, utilizing the proposed design specifications of ET-D~\cite{Hild_2011} and LISA L3~\cite{amaroseoane2017laser} as our detectors. We then use IMRPhenomXAS~\cite{PhysRevD.102.064001} to generate our source waveforms. If we assume all detected mergers are of PBH origin, then our minimum testable constraint is set for values of $f_\mathrm{PBH}$ and $\lra{M}$ where $N_\mathrm{det} \gtrsim 1$. We set our simulated observation period to $t_\mathrm{obs} = 1\,\mathrm{yr}$. 

\begin{figure}
    \centering
    \includegraphics[width=\linewidth]{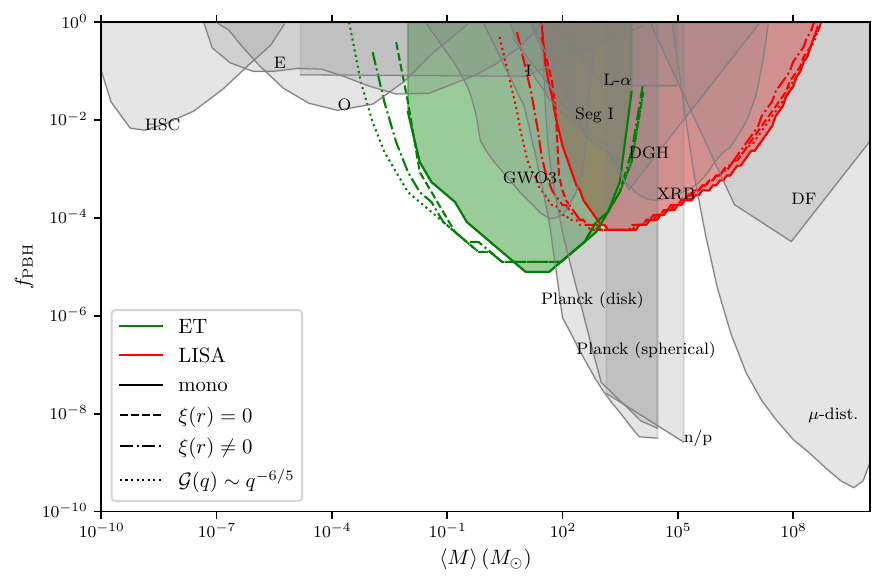}\\

    \includegraphics[width=\linewidth]{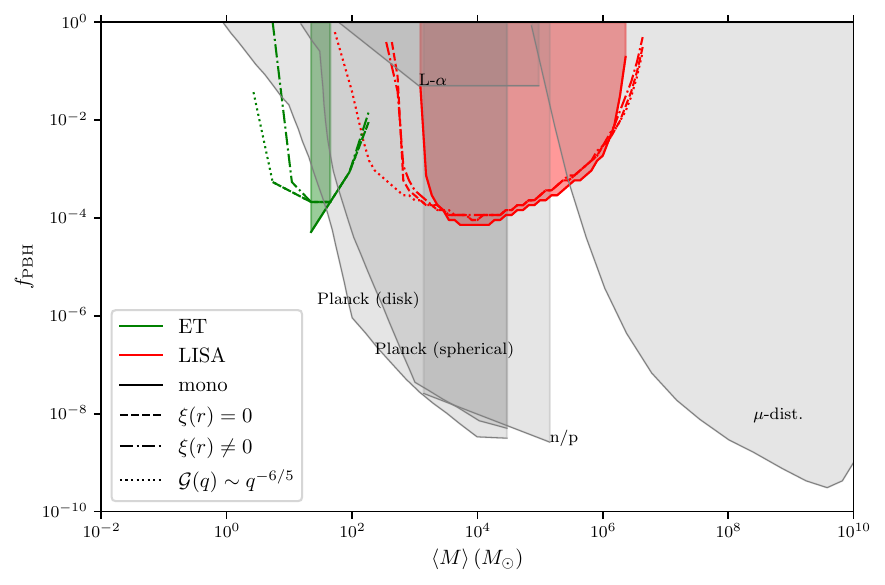}
    \caption{Projected constraints on PBH abundance $f_\mathrm{PBH}$ as a function of the average PBH mass $\lra{M}$ from simulated resolvable mergers at ET (green lines) and LISA (red lines) for monochromatic and $\gamma = -1/2$ power law distributions. Top panel shows constraints for merger events occurring in the redshift range $0.01 \leq z < 30$, while bottom panel shows constraints for mergers occurring in the redshift range $30 < z \leq 300$. The solid lines represent the constraints assuming monochromatic mass distributions, while the dashed and dash-dotted lines represent constraints for Poisson power law and correlated power law cases, respectively. The dotted lines represent the case where $\mathcal{G}(q)$ is set to $q^{-6/5}$. The gray regions are monochromatic constraints from other observations (see main text for the corresponding legend).}
    \label{fig:resolvable}
\end{figure}

Figure~\ref{fig:resolvable} shows the projected constraints from resolvable simulated mergers in ET (in green) and LISA (in red), assuming all resolved detections come from PBH mergers. The solid colored curves show the constraints if the mass distributions are assumed to be monochromatic, while the dashed and dash-dotted curves assume power law distributions with $\gamma = -1/2$. The dashed lines represent $N(y)$ for a purely Poissonian spatial distribution, while the dash-dotted lines represent PBHs with a correlated distribution. The dotted lines represent the case where we assume a non-zero correlation, but $\mathcal{G}(q)$ is set to $q^{-6/5}$. The top panel colored constraints are calculated from expected event rates $N_\mathrm{det}$ integrated from redshifts $0.01 \leq z < 30$, while the bottom panel constraints are calculated from events simulated to occur from redshifts $30 \leq z < 300$.

Two points are immediately clear from the top panel of Figure~\ref{fig:resolvable}, corresponding to low-redshift mergers. The first is that constraints from the correlated cases have their constraint window expanded toward the lower masses. The second is that there is practically no change to the high mass end of the constraint window, relative to the uncorrelated Poissonian case. We attribute the former to an enhancement in the differential merger rate from binaries with mass ratios $q \gg 1$. Lighter PBH distributions benefit from this enhancement the most, particularly ones with $\lra{M}$ smaller than $M_\mathrm{tot}$ detectors are most sensitive to. This is because high-$q$ binaries sampling one mass from the heavier end of the distribution gain both the reduced suppression \textit{and} higher detector SNR due to being closer to the peak detectable mass. Note however that this effect starts to taper off as we approach the minimum of the constraint curve. Past the minimum, the constraints from both correlated and uncorrelated cases are practically indistinguishable. Considering that high-$q$ binaries in the correlated case are always much less suppressed regardless of their total mass, this suggests that other factors are balancing out the enhancement. Some of these include the domination of the $M_\mathrm{tot}^{-32/37}$ factor in the merger rate, and the decrease in SNR of heavy binaries past the peak detection mass.

The bottom panel of Figure~\ref{fig:resolvable}, showing constraints from high-redshift mergers, paints a somewhat different story. The constraints from both Poissonian and standard (dash-dotted) correlated cases are practically indistinguishable, save for the low mass end of the constraint window, where this is a slight contraction of the constraint window for the ET case. At high redshifts, the detector SNR mass ratios $q \sim 10$ and above fall below the target threshold. Thus, even if the differential merger rate of high-$q$ binaries is boosted, a low value of detectability of the individual binaries would balance out the enhancement at the detection stage. The SNR for $q \gtrsim 1$ binaries would still be above threshold and provide the strongest signal at these redshifts. However, if these binaries are over-suppressed (as in the dash-dotted $\xi(r) \neq 0$ case) then their contribution to the merger rate and overall detection rate is significantly cut. If we normalize $\mathcal{G}(q) = 1$ when $q = 1$ (as in the dotted $\mathcal{G}(q) \sim q^{-6/5}$ case), then we get the usual low mass window extension. This implies that the $q \gtrsim 1$ merger channel provides a greater boost to the detection rate compared to much more unequal merging binaries, especially at higher redshifts.

\subsubsection{SGWB mergers}
For a merging population of PBHs with component masses $M_1$ and $M_2$, the expression for the expected SGWB spectrum at frequency $\nu$ is given by~\cite{phinney2001practicaltheoremgravitationalwave, Zhu_2011}
\begin{align}
\label{eq:omega}
    \Omega_\mathrm{GW}(\nu) = \frac{\nu}{\rho_0} \int_0^{z_\mathrm{cut}}\,\mathrm{d}z\,\mathrm{d}M_1\mathrm{d}M_2\,\frac{1}{(1+z)H(z)}\times\nonumber\\
    \times\frac{\mathrm{d}^2R_\mathrm{PBH}}{\mathrm{d}M_1\mathrm{d}M_2}\frac{\mathrm{d}E_\mathrm{GW}(\nu_S)}{\mathrm{d}\nu_S},
\end{align}
where the present energy density $\rho_0 = 3H_0^2 c^2 / 8\pi G$, $H_0$ is the Hubble constant, $H(z)$ is the Hubble parameter at redshift $z$, $R_\mathrm{PBH}$ is the PBH differential merger rate, and $E_\mathrm{GW}$ is the energy spectrum of GWs from the binary black hole (BBH) mergers evaluated at the redshifted source frequency $\nu_S = \nu(1+z)$. We assume a $\Lambda$CDM cosmology with $h = 0.7$, $\Omega_m = 0.3$, and $\Omega_\mathrm{de} = 0.7$. Note that the redshift integration is done up to $z_\mathrm{cut} = (\nu_\mathrm{cut}/\nu) - 1$, where $\nu_\mathrm{cut}$ is the cutoff frequency determined by the GW energy spectrum model. We utilize a form of the GW energy spectrum $E_\mathrm{GW}$ outlined in Ref.~\cite{PhysRevLett.106.241101}, taking the non-spinning limit. We calculate for the detection SNR $\rho_\mathrm{det}$ for the SGWB following Ref.~\cite{PhysRevD.88.124032}, setting the simulated observation period at $t_\mathrm{obs} = 1\,\mathrm{yr}$. We count a detection if $\rho_\mathrm{det}$ exceeds $\rho_\mathrm{thr} = 5$. 

\begin{figure}
    \centering
    \includegraphics[width=\linewidth]{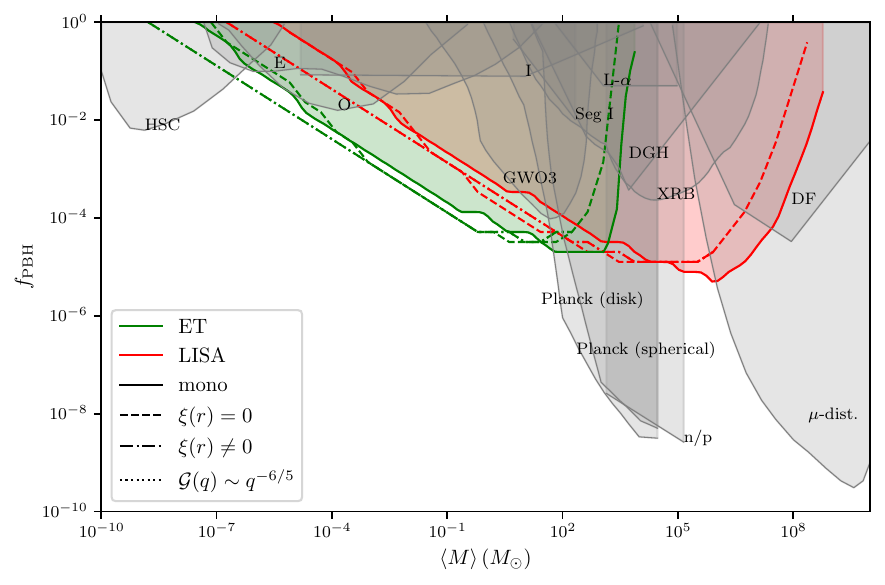}
    \caption{Projected constraints on PBH abundance $f_\mathrm{PBH}$ as a function of the average PBH mass $\lra{M}$ from simulated SGWB mergers at ET (green lines) and LISA (red lines) for monochromatic and $\gamma = -1/2$ power law distributions. The solid lines represent the constraints assuming monochromatic mass distributions, while the dashed and dash-dotted lines represent constraints for Poisson power law and correlated power law cases, respectively. The dotted lines represent the case where $\mathcal{G}(q)$ is set to $q^{-6/5}$. The gray regions are monochromatic constraints from other observations (see main text for the corresponding legend).}
    \label{fig:SGWB}
\end{figure}

Figure~\ref{fig:SGWB} shows the projected constraints from simulated SGWB-generating mergers in ET (in green) and LISA (in red), assuming the entire SGWB is generated by just PBH binary mergers. Just as above, the solid colored curves show the constraints if the mass distributions are assumed to be monochromatic, while the dashed and dash-dotted curves assume power law distributions with $\gamma = 0$. The dashed lines represent $N(y)$ for a purely Poissonian spatial distribution, while the dash-dotted lines represent PBHs with a correlated distribution. The dotted lines represent the case where we assume a non-zero correlation, but $\mathcal{G}(q)$ is set to $q^{-6/5}$.

In contrast to the constraints derived from resolvable mergers, which may see a degree of mass window contraction depending on the level of suppression of $q \gtrsim 1$ binaries, spatial correlation generally has a much more consistent impact on SGWB-derived constraints. We observe the same general trends as in the resolvable low-redshift mergers: an extension of the constraint window into the lower masses, and no change in the constraint window past the mass of the minimum testable constraint. Unlike the low-redshift constraints, which extend at most by two orders of magnitude, SGWB constraints have their constraint mass windows extended only up to half an order of magnitude at most. Relative to the Poissonian case, the correlated LISA constraint window shifts its lower window from $\lra{M}\sim 7.9\cdot10^{-6}\,M_\odot$ to $\lra{M} \sim 1.8\cdot10^{-7}\, M_\odot$, while the ET window extends from $\lra{M}\sim 7.2\cdot10^{-8}\,M_\odot$ to $\lra{M} \sim 1.7\cdot10^{-9}\, M_\odot$. In regions with expanded constraint, the testable abundance drops by at most half an order of magnitude. Of notable interest is the Einstein Telescope, as a non-detection of SGWB consistent with PBH mergers from ET may present the most stringent constraint yet for PBHs with average masses in the range $10^{-7}\,M_\odot < \lra{M} < 1\,M_\odot$, i.e. subsolar to planetary mass. We must point out however that our projections, even with the assumption of correlation, will not be able to reliably constrain the mass window below $\lra{M} < 10^{-7}\,M_\odot$ any more than what either existing HSC observations or evaporation constraints~\cite{PhysRevD.81.104019} have ruled out, even for extended mass distributions~\cite{PhysRevD.95.083508, PhysRevD.96.023514, PhysRevD.96.063507}. We also point out that the $\mathcal{G}(q) \sim q^{-6/5}$ constraints are indistinguishable from the constraints from standard correlation, suggesting that the $q\gtrsim 1$ merger channel alone does not impact SGWB constraints as much compared to the resolvable merger constraints. Nevertheless, both ET and LISA may still provide an avenue to test various assumptions about the PBH scenario, particularly initial spatial correlation, via measurements of the SGWB.

\section{\label{sec:conclusions}Conclusions and Limitations}
In this work, we present a modified form of the expected local PBH count $N(y)$ which accounts for possible underlying correlations in the density fluctuation field, entering through the function $\mathcal{G}(q)$ in Eq.~\ref{eq:mass_ratio_factor}. Our results show that binaries with an uneven mass ratio tend to have lower values of $N(y)$, leading to much lower possible suppression when set in a binary merger configuration. This can lead to much lower testable constraints on subsolar and planet-mass PBHs from ET and LISA, at least through potential observations of the SGWB.

It is worth pointing out a few caveats and limitations to our local count model. On its own, the correlation model~\cite{PhysRevD.109.123538} we use actually exhibits "zero-lag" behavior for binary mass ratios $q\sim 1$ as we increase $\sigma_r$ (equivalently, decrease the mass scale); this explains the over-suppression. We mitigate the divergent behaviour by forcefully setting the correlation to $\xi(r) = -1$ at the exclusion scale, although this does somewhat feel artificial. However, we do suspect that even in the case of a more realistic anti-correlation at short scales, $q\sim 1$ binary systems would still have at least the same order of magnitude of $N(y)$ as in the pure Poissonian case. If this is indeed true, then we expect the mass ratio dependence to behave similar to the $\mathcal{G}(q) \sim q^{-6/5}$ case near $q \gtrsim 1$. Note however that, strictly speaking, the above power law behaviour only applies to a specific set of parameters. These are a PBH power spectrum tilt of $n_p \simeq 0$ (or a mass distribution tilt of $\gamma \simeq -1/2$), a collapse threshold of $\delta_c \simeq 0.56$, and a mass ratio range of $2 \lesssim q \lesssim 1000$. Thus significant deviations from these parameters may also result in equally significant changes in how the mass ratio dependence behaves, as demonstrated in Figure~\ref{fig:g-q} for the case of $\delta_c = 0.10$.

Another curious point to our result is that our merger rate is not enhanced by positive clustering, but by \textit{negative} clustering due to the dominant anti-correlation effects imposed upon high mass ratio binaries. We argue this is only the case because of our choice of merger rate, which penalizes binaries in overdense clusters and promotes binaries in underdense regions. We do not explore it in this work, but a different choice of merger rate, such as one that accounts for compact three-body interactions~\cite{PhysRevD.101.043015, Raidal_2017, Raidal2025}, may actually benefit more from positive clustering. In particular, merger rates that scale with $N^{53/37}$ will see a slight enhancement in their merger rate for binary mass ratios $q\sim O(1)$, although in exchange higher mass ratios would have significantly lower merger rates. In this case, we expect a result similar to the constraints reported in Ref.~\cite{De_Luca_2021}. We must recognize however that existing three-body formulations of the early-time PBH merger rate are only valid for sufficiently narrow mass functions where the perturbing PBH $M_3$ is of similar order of mass compared to the mass of the binary $M_\mathrm{tot}$. Formulating a three-body merger rate for broad PBH mass distributions must necessarily and separately account for cases where $M_3 > M_\mathrm{tot}$ or $M_3 < M_\mathrm{tot}$.

To end things, let us discuss possible avenues for extension. Apart from the alternative merger rate models mentioned above, one can also explore the effects differently shaped power spectra $\mathcal{P}_\delta(k)$ may have on the local count and merger rate. In particular, spectra from various inflaton and quantum diffusion models~\cite{PhysRevD.96.063507, Pattison_2017, Biagetti_2018, Animali_2024} may introduce multi-modal peaks in the mass distribution, affecting clustering behavior~\cite{Animali_2024} and GW constraints. Introducing a variable collapse threshold $\delta_c(\sigma_r)$ may also be interesting, as Ref.~\cite{PhysRevD.109.123538} includes the possibility of it in their correlation model. On the low-redshift detection front, our results would be improved by accounting for foreground astrophysical signals and spin dynamics. The latter would necessitate the incorporation of accretion into the model, while we expect the former to only affect the constraints for masses above $M_\mathrm{PBH} \sim 1\,M_\odot$. Finally, we remark that GW constraints are not the only constraint class that can change if we involve clustering. Constraints from lensing observations may be impacted~\cite{Belotsky2019} if light, same-mass PBHs happen to form in close clusters. Similarly, the clustering evolution of PBHs~\cite{Luca_2020_2} may have to be re-examined for the case of PBH binaries with $q > 1$. The possible dynamics and orbital disruption of severely unequal mass ratio PBH binary systems needs to be explored in detail, especially once late-time structure formation effects are incorporated into the model.

\begin{acknowledgments}
    I acknowledge the use of Python package \texttt{gwent}~\cite{Kaiser:2021} in the generation of simulated detector SNR data for Einstein Telescope and LISA. I would also like to thank the referees for their insightful comments. This work was initiated while the author was affiliated with the National Institute of Physics at the University of the Philippines Diliman.
\end{acknowledgments}


\bibliography{apssamp}

\end{document}